\documentclass[a4paper,11pt]{article}
\title{Parton density functions for future electron-positron colliders}

\author{Andrej Arbuzov$^{a,b}$, Cecilia Romero$^{a,c}$,
Uliana Voznaya$^{a,b}$}

\begin{document}

\maketitle




\begin{center} {\it
$^a$ Bogoliubov Laboratory of Theoretical Physics,
Joint Institute for Nuclear Research,
Dubna, Moscow 141980, Russia \\
$^b$ Institute of Physics and Engineering, Dubna State University,
Universitetskaya str., 19, Dubna, Russia \\
$^c$ Nuclear Physics Department, Higher Institute of Technologies and Applied Sciences,
Salvador Allende ave., Havana, Cuba 
}
\end{center}


\begin{abstract}
Precision QED radiative corrections are critical for the physics programs of future high-energy $e^+e^-$ colliders. We review the state-of-the-art QED parton density function (PDF) formalism for resumming collinear photon and pair emissions. Our numerical analysis of initial-state radiation (ISR) highlights the slow convergence of the perturbative series and the growing significance of pair production at higher orders. Finally, we emphasize the need to implement these theoretical refinements into modern Monte Carlo generators to meet the stringent precision demands of upcoming experimental measurements.
\end{abstract}

\section{Introduction}
The next generation of electron-positron colliders, including FCC-ee, CEPC, and ILC are designed to operate at the absolute precision frontier of particle physics. To fully exploit their physics potential, theoretical predictions must match the experimental accuracy. A primary obstacle here is the treatment of QED radiative corrections. In high-energy scattering processes, the emission of collinear photons by the incoming and outgoing charged leptons generates terms proportional to the so-called large logarithm $L=\ln(Q^2/m_l^2)$ 
where $\sqrt{Q^2}$ is the characteristic energy scale of the hard process and $m_l$ is the lepton mass. At the high center-of-mass energies characteristic of these colliders, the actual perturbative expansion parameter $(\alpha L/(2\pi))$ renders fixed-order calculations insufficient and necessitates the calculation of higher-order contributions enhanced by these large logarithms.

The most robust and widely adopted framework for addressing this challenge is the electron Structure Function (SF) method. This approach treats the physical electron as a "composite" object, described by electron parton density (also called parton distribution) functions (PDFs) that describe the probability of finding an electron, positron, or photon carrying a specific momentum fraction of the initial particle. The formalism adapts the Dokshitzer-Gribov-Lipatov-Altarelli-Parisi (DGLAP) evolution equations, originally developed for QCD\footnote{Actually, the evolution equations were first written for QED~\cite{Lipatov:1971yp}.} to the abelian gauge theory of QED. The method was pioneered for $e^+e^-$ annihilation in the 1980s through the work of Kuraev and Fadin~\cite{Kuraev:1985hb}. Over the past decades, the formalism has evolved significantly, progressing from Leading Logarithmic (LL) approximation to more sophisticated Next-to-Leading (NLL) and Next-to-Next-to-Leading Logarithmic (NNLL) accuracies. Many modern implementations match the all-order LL resummation with exact matrix elements within Monte Carlo generators (such as BabaYaga~\cite{CarloniCalame:2003yt} and KKMC~\cite{Arbuzov:2020coe}) to accurately describe both the collinear/soft and hard/wide-angle photon emission regimes. 

The electron PDFs formalism is a prerequisite for the physics program of current and future lepton colliders. First, it is essential for the precise determination of the collider luminosity via small and large angle Bhabha scattering, which must be known up to $10^{-5}$ accuracy~\cite{Jadach:2019bye} to serve as a reliable normalization for all other cross-section measurements. Second, at the $Z$ and $W$ poles, the extraction of high-precision electroweak observables (the effective weak mixing angle, the $Z$ boson mass, and the $W$ boson width) requires the rigorous resummation of QED initial and final state radiation. Without precise PDFs, residual theoretical uncertainties could obscure subtle Standard Model deviations or mask signals of New Physics~\cite{dEnterria:2016fpc}. Finally, for multi-TeV machines like CLIC or the high-energy stage of FCC-ee, the formalism must be further extended to resum electroweak Sudakov logarithms, ensuring that theoretical uncertainties do not become the limiting factor in our search for new physics. The method is applicable not only to collider processes. 
In these proceedings, we review the current state-of-the-art in QED PDFs and discuss the necessary refinements and extensions required to meet the stringent theoretical demands of the future electron-positron collider era.

\section{Structure function formalism in QED}
In the standard QCD factorization framework, inclusive cross sections decompose into universal, long-distance PDFs and perturbative, short-distance partonic cross sections~\cite{Collins:2004nx}. A parallel approach applies to QED, where collinear radiation of photons and light pairs from incoming leptons is resummed into QED PDFs. Since these functions factorize at the hard subprocess, we refer to them as "QED PDFs" to maintain consistency with QCD notation, rather than using the traditional term "QED structure functions".

Consider the distribution of the system produced in $e^+e^-$ annihilation with respect to its invariant mass squared, $s'$. This distribution can be expressed in a form analogous to that of the Drell-Yan process: 
\begin{equation}
\frac{d\sigma_{e^+e^-\to\gamma^*,Z}}{ds'}=\sum_{a,b=e^-,\gamma,e^+}D_{ae^-}\otimes D_{be^+}\otimes\frac{d\hat\sigma_{ab\to\gamma^*,Z}}{ds'},
\label{eq:fact}
\end{equation}
with the perturbative partonic cross section
\begin{equation} \nonumber
\hat\sigma_{ab}=\sigma^{(0)}_{ab}+\frac{\alpha}{2\pi}\bar\sigma^{(1)}_{ab}+\Big(\frac{\alpha}{2\pi}\Big)^2\bar\sigma^{(2)}_{ab}+\dots,
\end{equation}
where bars denote subtraction of mass singularities in the $\overline{\rm MS}$ scheme. Here $s$ is the $e^+e^-$ c.m.\ energy squared, $a,b$ label the partons entering the hard process ($e^-$, $e^+$, $\gamma$), and $D_{ae^-}$, $D_{be^+}$ are their longitudinal momentum distributions with momentum fractions $z_{1,2}$. 

At NLO, the QED evolution equations in integral form are~\cite{Arbuzov:2022fmv}
\begin{equation}
D_{ba}\!\left(x,\frac{\mu_F^2}{\mu_R^2}\right)=\delta(1-x)\,\delta_{ba}+\frac{\alpha}{2\pi}d^{(1)}_{ba}\!\left(x,\frac{\mu_R^2}{m_e^2}\right)
+\!\sum_{c=e,\bar e,\gamma}\int_{\mu_R^2}^{\mu_F^2}\frac{dt\,\alpha(t)}{2\pi t}\int_x^1\!\frac{dy}{y} P_{bc}\left(\frac{x}{y}\right) D_{ca}\!\left(y,\frac{t}{\mu_R^2}\right).
\label{eq:evol}
\end{equation}
$D_{ba}$ is the probability of finding parton $b$ with momentum fraction $x$ in the initial particle $a$; $b,c\in\{e,\bar e,\gamma\}$. The variable $y$ is the fraction carried by the intermediate parton $c$, so $x/y$ is the fraction passed from $c$ to $b$. $P_{bc}$ are the splitting functions, $t$ runs from $\mu_R^2$ to $\mu_F^2$ with running coupling $\alpha(t)$, $\delta(1-x)\delta_{ba}$ is the no-emission term, and $d^{(1)}_{ba}$ is the $\overline{\rm MS}$ initial condition. Parameters $\mu_F$ and $\mu_R$ are the factorization and renormalization scales. Note that beyond the Born level, $\hat\sigma$ also depends on $\mu_F$.

The splitting functions are
\begin{equation}
P_{bc}(x,t)=P^{(0)}_{bc}(x)+\frac{\alpha(t)}{2\pi}P^{(1)}_{bc}(x)+\mathcal{O}(\alpha^2).
\label{eq:split}
\end{equation}
Both the splitting functions and initial conditions are obtained by taking the Abelian limit of the QCD results. Furthermore, the QED fragmentation functions are linked to the PDFs through the Gribov-Lipatov relations.

The factorization scale $\mu_F$ separates the collinear radiation resummed into the PDFs from the hard interaction, while $\mu_R$ acts as the renormalization scale for $\alpha$. Although physical observables must be scale-independent to all orders in perturbation theory, their residual dependence on $\mu_F$ and $\mu_R$ at finite order serves as an estimate of uncalculated higher-order contributions. Furthermore, the factorization scheme prescribes how collinear terms are partitioned between the PDFs and the partonic cross section $\hat{\sigma}$; while individual components are scheme-dependent, the physical result is not. The $\mu_F$ dependence of ISR corrections in $e^+e^-$ annihilation is detailed in Ref.~\cite{Arbuzov:2025pui}.

In recent years, explicit analytic results for QED PDFs at the next-to-leading logarithmic approximation have been independently derived and partially cross-checked in Refs.~\cite{Arbuzov:2022fmv,Blumlein:2011mi,Frixione:2012wtz}. A significant step toward the NNLL approximation was recently achieved: Refs.~\cite{Stahlhofen:2025hqd,Schnubel:2025ejl} presented the complete analytic results for $\mathcal{O}(\alpha^2)$ QED PDFs.

The singular behavior of QED PDFs as $x \to 1$ is dictated by the endpoint structure of the splitting functions and the initial conditions, and arises from infrared singularities associated with soft photon or light-pair emission. These singularities are regularized through cancellations with corresponding virtual loop contributions, as guaranteed by the Bloch-Nordsieck theorem. As shown in the next section, the convergence of the perturbation series is relatively slow, making higher-order contributions significant. In such cases, a partial resummation of the series is warranted. Resummation within the LL approximation was previously well established~\cite{Kuraev:1985hb,Cacciari:1992pz,Passarino:2001wy}. 
Ref.~\cite{Arbuzov:2026qpr} presents a modified factorization scheme for the simultaneous exponentiation of pure photonic and non-singlet pair corrections, which is then matched to existing higher-order analytic calculations.

\section{Numerical results}
In Table 1, we present the numerical results for the initial state radiation (ISR) corrections to the cross section of electron-positron annihilation. The results are expressed as percentages relative to the Born cross section, calculated at $\sqrt{s} = M_Z \approx 91.1876$ GeV and $\sqrt{s} = 240$ GeV, with energy cuts $z_{min} = 0.3$ and $0.8$. The factorization scale is set to $\mu_F = \sqrt{s}$. The individual contributions at different orders in $\alpha$ and $L$ are defined via the expansion
\begin{equation}
\frac{d\sigma_{\bar{e}e}^{\mathrm{NNLO}}(s')}{ds'} = \frac{\sigma_{\bar{e}e}^{(0)}(s')}{s} \left\{ \delta(1-z) + \sum_{\stackrel{k=1}{k \geq l \geq k-2}}^{\infty} c_{kl}(z) \left( \frac{\alpha}{2\pi} \right)^k L^l + \mathcal{O}(\alpha^k L^{k-3}) \right\},
\end{equation}
where $z = z_1z_2$ and coefficients $c_{ij}(z)$ are taken from Ref.~\cite{Arbuzov:2024tac}.
Separate contributions to the ISR corrections in different orders in $\alpha$ and $L$ are defined as
\begin{eqnarray}
  h_{ij} = \left(\ \, \int\limits_{z_{min}}^{1}dz \left(\frac{\alpha}{2\pi}\right)^{i}L^jc_{ij}(z) \right)\cdot 100\%,
\end{eqnarray}
where $z_{min}$ corresponds to the lower experimental limit on the invariant mass of the final $\mu^+\mu^-$ pair: $\sqrt{s'}>\sqrt{sz_{min}}$.

\begin{table}[ht] 
	\caption{}
	\begin{tabular}{|l|c|c|c|c|c|c|c|c|c|c|} 
		\hline  
		& $ h_{11}$ & $ h_{10}$ & $h_{22}$ & $h_{21}$ & $h_{20}$ & $h_{33}$ &$h_{32}$ & $h_{44}$ & $h_{43}$ & $h_{55}$  \\
		\hline
		\multicolumn{11}{|c|}{$\sqrt{s}=M_Z$}\\
		\hline
		\multicolumn{11}{|c|}{$z_{min}=0.3$}\\
		\hline 
		$\gamma$ & -32.74674
		& 2.00210 & 4.88350
		& -0.59513 & 0.01521
		& -0.37757 & 0.07102 & 0.00345 & -0.00185 & 0.00315 \\
		\hline
		Pairs & 0  & 0  & -0.30633
		& 0.15852 & -0.02151
		& 0.08757 & -0.04604 & -0.00909 & 0.00375 & -0.00008  \\
		\hline
		Full &-32.74674
		& 2.00210
		& 4.57747
		& -0.43662 & -0.00630
		& -0.29002 & 0.02499 & -0.00564 & 0.00190 & 0.00307 \\
		\hline 
		\multicolumn{11}{|c|}{$z_{min}=0.8$}\\
		\hline 
		$\gamma$ & -32.82241
		& 2.00521  & 4.88622
		& -0.59589 & 0.01524
		& -0.37697 & 0.07094 & 0.00338 & -0.00184 & 0.00315\\
		\hline
		Pairs & 0& 0  & -0.30719
		& 0.15883 & -0.02150
		& 0.08763 & -0.04604 & 0.00907 & 0.00374 & -0.00008 \\
		\hline
		Full & -32.82241
		& 2.00521
		& 4.57904
		& -0.43706 & -0.00626
		& -0.28984 & 0.02490 & -0.00568 & 0.00190 & 0.00307\\
		\hline
		\hline
		\multicolumn{11}{|c|}{$\sqrt{s}=240$ GeV}\\
		\hline 
		\multicolumn{11}{|c|}{$z_{min}=0.3$}\\
		\hline 
		$\gamma$ & 11.88597 &  0.19300 & -0.66463 & 0.16592
		& -0.00779 &  -0.03637
		& 0.00367 & 0.00250 & -0.00120 & $2 \cdot 10^{-7}$\\
		\hline
		Pairs & 0 & 0
		&  0.23207
		& -0.07145 & -0.00710
		&  -0.01958& 0.00776 & -0.00126 & 0.00087 & 0.00009\\
		\hline
		Full & 11.88597 &  0.19300 &  0.43256
		&  0.09446 & -0.01480
		& -0.05595 & 0.01143 & 0.00124 &  -0.00033 & 0.00009 \\
		\hline
		\multicolumn{11}{|c|}{$z_{min}=0.8$}\\
		\hline 
		{$\gamma$} & -19.71071 & 1.02535  & -0.64157
		& 0.01773 & -0.00046
		& 0.15716 & -0.02851 & -0.00960 & 0.00267 & 0.00002\\
		\hline
		Pairs &  0 & 0
		& -0.09742
		& 0.02401 &  -0.00297
		& -0.01480 & 0.00995 & 0.00455 & -0.00250 & -0.00028\\
		\hline
		Full & -19.71071 & 1.02535  & -0.73989
		& 0.04174 & -0.00343
		& 0.14236 & -0.01856 & -0.00505 & 0.00018 & -0.00026\\
		\hline 
	\end{tabular}
\label{t0308} 
\end{table}

One can see that the series in $\alpha^n L^n$ converges slowly, particularly in the soft region $x \to 1$, which dominates the cross section at the $Z$-resonance.  This slow convergence underscores the necessity for resummation, or exponentiation, of these radiative corrections to achieve the precision required for future lepton colliders~\cite{Arbuzov:2026qpr}.
Note that pair production contributions, which appear starting at $O(\alpha^2 L^2)$, provide partial compensation to the pure photonic terms. For example, at the $Z$-peak, the photonic term $h_{22}^{\gamma} \approx 4.88$ is partially offset by the pair contribution $h_{22}^{\mathrm{pairs}} \approx -0.31$. Furthermore, the relative weight of pair corrections grows with the order; they constitute approximately 25\% of the photonic term at $\mathcal{O}(\alpha^3 L^3)$ and exceed it at $\mathcal{O}(\alpha^4 L^4)$.
At $\sqrt{s} = M_Z$, the resonance behavior suppresses hard radiation, resulting in results that are nearly independent of the $z_{min}$ cut. In contrast, at $\sqrt{s} = 240$~GeV, the dependence on the energy cut is pronounced. 
At the $Z$-resonance, we observe an alternating sign pattern in the leading terms $h_{11}, h_{22}$, and $h_{33}$. However, this pattern does not persist in the higher-order terms, indicating a complex structure in the convergence of the perturbative series.

\section{Conclusions}
The QED PDF formalism remains a cornerstone for achieving the theoretical precision required by next-generation $e^+e^-$ colliders. Recent theoretical developments have significantly advanced this framework, pushing accuracy from leading-logarithmic approximations toward NNLL and effectively addressing the complexities of collinear photon and pair emission. The universality of this method makes it an essential tool for the experimental physics program, providing the rigor necessary for luminosity determination, precision electroweak measurements, and the search for subtle signals of New Physics.

Looking ahead, the transition from analytic refinement to practical application is critical. While the current state-of-the-art results for QED PDFs are robust, the ability to fully exploit them in experimental data analysis depends on their integration into software tools. Future work must be dedicated to the systematic implementation of these higher-order results into computer codes and modern Monte Carlo generators. Bridging this gap between theoretical development and computational availability is a prerequisite for ensuring that theoretical uncertainties do not become the limiting factor at upcoming facilities such as FCC-ee, CEPC, and ILC.

\bibliographystyle{jhep}  
\bibliography{arbuzov_ICHEP_2026}

%
%

\end{document}